# ALIASNET: ALIAS ARTEFACT SUPPRESSION NETWORK FOR ACCELERATED PHASE-ENCODE MRI

## AUTHORS

Mr. Marlon Bran Lorenzana[1]

Dr. Shekhar S. Chandra Ph.D[1]

Prof. Feng Liu[1]

[1]School of Information Technology and Electrical Engineering, University of Queensland, Australia

## CORRESPONDING AUTHOR

Marlon Bran Lorenzana

School of Information Technology and Electrical Engineering, University of Queensland, St. Lucia, Queensland, 4072, Australia.

Email: uqmbran@uq.edu.au

# Abstract

Sparse reconstruction is an important aspect of MRI, helping to reduce acquisition time and improve spatial-temporal resolution. Popular methods are based mostly on compressed sensing (CS), which relies on the random sampling of k-space to produce incoherent (noise-like) artefacts. Due to hardware constraints, 1D Cartesian phase-encode under-sampling schemes are popular for 2D CS-MRI. However, 1D under-sampling limits 2D incoherence between measurements, yielding structured aliasing artefacts (ghosts) that may be difficult to remove assuming a 2D sparsity model. Reconstruction algorithms typically deploy direction-insensitive 2D regularisation for these direction-associated artefacts. Recognising that phase-encode artefacts can be separated into contiguous 1D signals, we develop two decoupling techniques that enable explicit 1D regularisation and leverage the excellent 1D incoherence characteristics. We also derive a combined 1D + 2D reconstruction technique that takes advantage of spatial relationships within the image. Experiments conducted on retrospectively under-sampled brain and knee data demonstrate that *c*ombination of the proposed 1D AliasNet modules with existing 2D deep learned (DL) recovery techniques leads to an improvement in image quality. We also find AliasNet enables a superior scaling of performance compared to increasing the size of the original 2D network layers. AliasNet therefore improves the regularisation of aliasing artefacts arising from phase-encode under-sampling, by tailoring the network architecture to account for their expected appearance. The proposed 1D + 2D approach is compatible with any existing 2D DL recovery technique deployed for this application.

# 1 INTRODUCTION

The theory of Compressed Sensing (CS) [1,2] is integral to sparse image reconstruction and has seen application in many areas of signal processing [3]. It is an especially important technology for magnetic resonance imaging (MRI), which is an intrinsically lengthy process subject to physical and physiological constraints. Such constraints necessitate the sequential acquisition of k-space (discrete Fourier space) through continuous trajectories [4,5]. Scan times are subsequently influenced by available sampling and reconstruction methods [6,7], where implementation of CS can improve spatial-temporal resolution and increase scanner availability. From a signal processing perspective, CS performs optimally when three conditions are met: incoherent under-sampling, transform sparsity and non-linear optimisation. Over the past decade, there has been active development of both suitably incoherent sampling schemes for MRI [8], and the algorithms deployed in reconstruction [9–11].

Incoherence of a sampling matrix is often synonymous with orthogonal measurement and adhering to the well-known restricted isometry property (RIP) [12,13]. The objective is to produce noise-like image artefacts that can be easily distinguished from image features. Ideal measurement is therefore non-deterministic or an approximation thereof, as random sampling has been shown to produce incoherence with high probability. Unfortunately for MR applications, pure Cartesian random sampling is impractical to implement as a two-dimensional (2D) acquisition sequence [9]. Instead, practical 2D Cartesian CS fully samples in the frequency-encode direction and under-samples in the phase-encode direction [4,5]. Incoherence between these measurements is thereby one-dimensional (1D), which yields structured aliasing artefacts (ghosts) along the under-sampled axis. Strategies have been developed to mitigate this impact on image quality, such as varying the density of measurement [5,14]. This is known as multi-level sampling, where acquisition is asymptotically coherent with the regularising function, ensuring that high energy regions of k-space are well captured [15,16]. For reconstruction algorithms, another approach suggests first recovering an image from the under-sampled 1D columns of k-space [17], fully exploiting randomness in the available direction.

Early reconstruction techniques utilised pre-defined sparsifying transforms for image regularisation, where under-sampling artefacts remain noise-like and image features could be extracted [4,18]. For successful implementation, these non-adaptive transforms rely on the sampling scheme to produce incoherent and unstructured artefacts. Adaptive methods instead learn the regularising function directly, ensuring a sparse representation throughout the reconstruction process [19–24]. However, general downsides to sparse encoding include the introduction of additional image transformations, assumptions of image appearance, and necessitating increased calculations per iteration of CS [25]; learned sparse encoders exacerbate additive computation cost. Other prior knowledge has also been incorporated into the recovery process, such as a low-rank constraint [26,27], or redundancies in multi-coil MRI [28–31]. Machine learning elevated this notion of learned regularisation by expressing CS-MRI solutions as deep neural networks (DNN). Their data-driven approach allows for task-specific, highly non-linear operations that consistently outperform conventional CS recovery techniques [11,32]. For the most part, deep learned (DL) methods operate in the image domain [33], with k-space measurements integrated into the loss function [34–36] or data consistency layer [37–46]. Alternatively, k-space interpolation networks [47,48] or direct k-space to image transformation [49,50] have also been proposed. As CS-MRI artefacts degrade an image in non-local and potentially unrecoverable ways, recent contributions have instead proposed a dual-domain reconstruction approach. These networks execute operations in k-space alongside image denoising layers, enabling recovery of image features that may be otherwise lost from single-domain techniques. Some explore k-space recovery in the denoising sense [51–55], and others perform k-space interpolation via k-space redundancies [56,57]. Although dual-domain networks excel at recovering images from random 2D Cartesian and projection-based under-sampling patterns, they typically deploy convolutional neural network (CNN) operations directly in k-space. Reconstruction is thereby limited to interpolation from known values, with large distances between sampled points necessitating elaborate network architectures and high computation costs to overcome. Additionally, regularisation is still often limited to an idealised 2D transformation, which may not be suitable for certain image artefacts. For example, Iterative Shrinkage-Thresholding Algorithm Network (ISTA-Net) [39] and Deep cascade of Convolutional Neural Networks (DcCNN) [37] assume that the image can be represented as sparse in the 2D transform (ISTA-Net) or 2D denoising (DcCNN) sense.

In this paper, we develop a CS methodology tailored to the recovery of 1D Cartesian random under-sampled MRI. Our method employs a series of 1D CS operations that efficiently populate missing k-space to improve the image estimate (by alleviating aliasing artefacts), before 2D regularisation. This regularisation is executed by 1D CNN modules, which are then combined with existing 2D DL reconstruction techniques. Other notable 1D DL formulations have been proposed, such as domain-transform manifold learning in phase-encoding direction [50] and One-dimensional Deep Low-rank and Sparse Network [57]. To the best of our knowledge, however, this is the first 1D only network designed to supplement existing 2D reconstruction techniques. Our contributions are summarised as follows:

- A general 1D CS framework for MRI that decouples under-sampled k-space into a series of 1D signals in two separate domains. Asymptotic incoherence between sampling and regularisation is improved by explicitly recovering missing data in the direction of under-sampling.

- Development of an efficient proximal mapping for DL non-linear regularisers in our proposed algorithm. This constitutes the proposed 1D CNN regularisation modules.
- The combination of our 1D modules with existing 2D DL recovery methods efficiently achieves image quality superior to simply increasing the size of the original 2D network.
- Comparisons with state-of-the-art dual-domain reconstruction techniques highlight that our CS-based recovery of missing k-space is better suited to phase-encode under-sampling MRI.

The proposed method is an **Alias** artefact suppression **Net**work (AliasNet) that enhances the regularisation of 1D under-sampled MRI. An overview of the approach is visualised in Figure 1. Relevant theory and the proposed method are described in Section 2. Experiments with a purely 1D reconstruction, as well as integrating the model into well-known networks DcCNN [37] and ISTA-Net [39] are investigated in Section 3. Finally, Section 4 contains a discussion of these experiments.

## 2 METHODS

### 2.1 COMPRESSED SENSING

One can model CS for MRI by considering an object MR image $x_0$ and the subset of associated k-space measurements $y$, such that $y = RFx_0 + \sigma$. Here $F$ represents the 2D discrete Fourier transform (DFT), $R$ is the under-sampling matrix and $\sigma$ is complex Gaussian noise. As equation $x = (RF)^H[y - \sigma]$ is ill-posed, image estimate $x$ can be recovered according to,

$$\arg\min_{\mathbf{x}} \|RF\mathbf{x} - \mathbf{y}\|_2^2 + \rho h(\mathbf{x}), \qquad (1)$$

where $h(\cdot)$ is a regularisation function. Commonly, $h(\cdot)$ is chosen such that $\rho h(x) = \rho\|\Psi(\mathbf{x})\|_1$, which enforces sparsity in some transform $\Psi(\cdot)$; $\rho$ is a weighting factor. As asymptotic incoherence between sampling and sparsifying functions is important for the successful implementation of CS [15,16], it can be beneficial to consider the limitations of available sampling schemes when selecting the regularisation employed. In this work, we explore the addition of explicit DL 1D regularisation for phase-encode under-sampled MRI.

#### 2.1.1 1D Vs 2D Artefacts

Figure 2 illustrates artefacts associated with 2D random Cartesian (a) and 1D random phase-encode (b) under-sampling schemes. The central images correspond to each sampling scheme's point spread function (PSF). Convolution between the PSF and the ground truth image yields the under-sampled image. Visualisation of a sampling mask's PSF thereby illustrates how pixels interfere during under-sampling. While the 2D strategy produces unstructured and noise-like image artefacts, it is not a practical sampling pattern for 2D acquisition schemes. On the other hand, we see 1D artefacts arise as aliasing "ghosts" of surrounding image features. This stems from their respective PSF, in which the 2D strategy resembles the 2D dirac delta function. Conversely, all non-zero coefficients of the 1D under-sampling are large in amplitude and located in the central column. Importantly, the 1D PSF resembles the 1D dirac delta function. As adjoining image columns are similar in appearance, the resulting 1D ghosts are also similarly structured. It is therefore difficult for reconstruction algorithms to identify and remove these structures using non-directional 2D operations. As such, we suggest image regularisation be performed in the direction of under-sampling via a 1D CNN approach, leveraging incoherence in the available direction.

## 2.2 DECOUPLING 2D K-SPACE INTO 1D SIGNALS

To leverage the excellent 1D incoherence characteristics of phase-encode under-sampling, we must decouple image $x$ and its k-space $y$ into contiguous 1D signals. We identify two possible formulations of the problem by considering the following equality,

$$y = Fx = \Phi_P \Phi_F x. \qquad (2)$$

$\Phi_F$ and $\Phi_P$ are the 1D DFT in the frequency- and phase-encode direction respectively. We may then relate columns of image $x$ and k-space $y$ with the following two equations,

$$(\Phi_F^H y)_i = (\Phi_P x)_i \qquad (3)$$

$$(\Phi_P^H y)_i = (\Phi_F x)_i, \qquad (4)$$

where $i$ denotes the extracted $i^{th}$ column. Figures 1c, 1d demonstrate this relationship.

### 2.2.1 Phase-Encode Alias Artefact Suppression (Image Domain)

Recognising that artefacts arising from phase-encode under-sampling can be characterised as 1D perturbations of Eq. (3), Yang et al. [17] provide an "upgraded" image estimate before 2D CS. Their proposed optimisation problem can be expressed as,

$$\min_{\mathbf{x}} \sum_i \left\| R_{1D}(\Phi_P x)_i - (\Phi_F^H y)_i \right\|_2^2 + \rho \mathbf{TV}_{1D}(x_i), \qquad (5)$$

$R_{1D}$ is the 1D under-sampling pattern and $\rho$ is a weighting factor. As visualised in Figure 1d, 1D columns $x_i$ are related to Fourier measurements $(\Phi_F^H y)_i$ via the 1D DFT. Therefore, recovery of missing k-space is performed by enforcing a total-variation (TV) constraint on columns $x_i$, whilst ensuring consistency with related intermediate Fourier measurements $(\Phi_F^H y)_i$.

### 2.2.2 Phase-Encode Alias Artefact Suppression (Intermediate Fourier Domain)

As image domain artefacts are often non-local and structured, dual-domain techniques perform convolutional operations directly in k-space [51–55]. However, large distances between sampled points may limit the utility of this approach. Therefore, alongside 1D image domain artefacts, we also consider artefacts arising in the intermediate Fourier domain described by Eq. (4). Our proposed optimisation problem is therefore,

$$\min_{\mathbf{x}} \sum_i \|R_{1D}(\Phi_P \Phi_F \mathbf{x})_i - \mathbf{y}_i\|_2^2 + \rho h_{1D}((\Phi_F \mathbf{x})_i | \Theta_P). \tag{6}$$

In this instance, k-space values will be populated by enforcing learned constraint $h_{1D}(\cdot | \Theta_P)$ on aliased 1D signals of intermediate domain $(\Phi_F \mathbf{x})_i$ (Figure 1c). For simplicity, we instead define,

$$G(\mathbf{u}, \mathbf{v} | \Theta) = \|R_{1D} \Phi_P \mathbf{u} - \mathbf{v}\|_2^2 + \rho h_{1D}(\mathbf{u} | \Theta), \tag{7}$$

which allows Eqs. (5, 6) to be written concisely for each method as $G(\mathbf{x}_i, (\Phi_F^H \mathbf{y})_i | \Theta_F)$ and $G((\Phi_F \mathbf{x})_i, \mathbf{y}_i | \Theta_P)$ respectively; $\Theta_F$ and $\Theta_P$ are the set of CNN parameters for image and intermediate Fourier domain regularisation. Expressing Eq. (7) in this manner highlights that arbitrary regularisers $h_{1D}(\cdot | \Theta)$, which are suitably incoherent with respect to $\mathbf{y}$ in image $\mathbf{x}$ or intermediate $\Phi_F \mathbf{x}$ domain can be deployed for image recovery.

## 2.3 IMPLEMENTATION

To efficiently suppress 1D artefacts, we propose to solve Eq. (7) via the proximal gradient (PG) descent technique described by Algorithm 1.

---

**Algorithm 1:** 1D Iterative Update

---

**Result:** Recover $\mathbf{u}_0$ from measurements $\mathbf{v}$.

**Function:** $\underset{\mathbf{u}}{\operatorname{argmin}} \|R_{1D} \Phi_P \mathbf{u} - \mathbf{v}\|_2^2 + \rho h_{1D}(\mathbf{u} | \Theta)$

**Parameters:**

- $n_t$ – Number of iterations.
- $\rho$ – Regularisation strength.
- $h_{1D}(\cdot | \Theta)$ – Indicator function of noiseless $\mathbf{u}$.
- $\Theta$ – Set of CNN weights.

**Init:** Set $t = 0$, $\mathbf{u}^0 = \Phi_P^H \mathbf{v}$

**While** $t < n_t$ **do**:

$$\mathbf{z}^t = R_{1D} \Phi_P \mathbf{u}^{t-1} - \mathbf{v} \tag{8}$$

$$\mathbf{r}^t = \mathbf{u}^{t-1} - \rho^t \Phi_P^H \mathbf{z} \tag{9}$$

$$\boldsymbol{u}^t = \underset{\boldsymbol{u}}{\mathrm{argmin}} \|\boldsymbol{u} - \boldsymbol{r}^t\|_2^2 + h_{1D}(\boldsymbol{u}|\Theta) \quad (10)$$

**end**

In this approach, the fully sampled 1D signal $\boldsymbol{u}_0$ is recovered from measurements $\boldsymbol{v}$ by enforcing $h_{1D}(\boldsymbol{u}|\Theta)$ in each iteration. If $h_{1D}(\cdot|\Theta)$ is an indicator function of set $C$, where $C$ is the set of noiseless MRI columns for which a family of denoisers $g(\cdot)$ exist, then Eq. (10) is a special case of proximal mapping where $\boldsymbol{u}^t = \mathrm{prox}_{h_{1D}}(\boldsymbol{r}^t)$, and therefore,

$$\boldsymbol{u}^t = \arg\min_{\mathbf{u} \in C} \|\boldsymbol{u} - \boldsymbol{r}^t\|_2^2 = g(\boldsymbol{r}^t). \quad (11)$$

This assumes that $\boldsymbol{r}^t = \boldsymbol{u}^{t-1} - \rho^t \Phi_P^H \boldsymbol{z}^t$ can be modelled as $\boldsymbol{r}^t = \boldsymbol{u}_0 + \boldsymbol{\epsilon}^t$, where $\boldsymbol{\epsilon}^t$ is residual under-sampling noise. Because the 1D under-sampling is incoherent with respect to $\boldsymbol{u}_0$ (see Figure 2b), it is expected that $\boldsymbol{\epsilon}^t$ can be modelled as signal independent and noise-like [17,57]. We, therefore, propose to apply $h_{1D}(\cdot|\Theta)$ via $g_{cnn}(\cdot|\theta_t)$, which allows for a data-driven mapping between noisy measurements $\boldsymbol{r}^t$ and fully sampled 1D signal $\boldsymbol{u}_0$. Eq. (10) can be written as,

$$\boldsymbol{u}^t = g_{cnn}(\boldsymbol{r}^t|\theta_t). \quad (12)$$

Here $\theta_t$ are the CNN parameters at iteration $t$. Figure 3a illustrates our chosen architecture for $g_{cnn}(\cdot|\theta_t)$, which features a skip connection for residual learning. As discussed in [39,48], residuals of natural images are highly compressible, consisting mainly of high-frequency components. Given our proposed $g_{cnn}(\cdot|\theta_t)$ is applied directly to $\boldsymbol{r}^t$, we deploy this residual architecture to encourage sparsity in the output signal. As the 1D regularisation is intended to complement existing 2D reconstruction algorithms, $g_{cnn}(\cdot|\theta_t)$ is composed of just 3 1D CNN operations with intermediate Leaky ReLU activation functions. The kernel size is 9 and the feature size is 8. Therefore, each iteration of Algorithm 1 requires just 885 parameters, yielding a lightweight 1D regularisation module. Figure 3b illustrates the update process outlined in Algorithm 1. It should be noted that the $g_{cnn}(\cdot|\theta_t)$ and $\rho^t$ learned for a given iteration is only applied to either intermediate Fourier or image space. This is because features of the two domains are expected to differ and therefore, domain-specific regularisation is more suitable. The following sections will define intermediate Fourier and image domain regularisation as $G_P$ and $G_F$ respectively; their iterations are defined as $n_p$ and $n_f$.

### 2.3.1 Combination with 2D Techniques

Given that structured and non-local artefacts are typical of phase-encode under-sampling (see Figure 2b), we propose a combined regularisation that directly penalises 1D perturbations via AliasNet. 2D transforms then leverage spatial relationships to further enhance image quality. Additionally, for each $N \times M$ image there will be $M$ 1D training samples available, ensuring additional parameters incurred by the 1D modules are well trained. We demonstrate superior scaling with a total number of parameters compared to simply increasing the size of the original 2D network. We also compare against Dual-Domain Recurrent Network (DuDoRNet) [53] in various tests. Despite being more computationally efficient, our combined 1D + 2D networks consistently outperform alternatives. In the general case, we define our combined 1D and 2D optimisation problem as follows,

$$\min_{\mathbf{x}} \|RF\mathbf{x} - \mathbf{y}\|_2^2 + \lambda \Psi_{2D}(\mathbf{x}) + G_P + G_F. \quad (13)$$

Where $G_P$ and $G_F$ are defined in Eqs. (14, 15) and penalise 1D perturbations, $\Psi_{2D}(\mathbf{x})$ is the 2D regularisation. The combined optimisation is solved via an alternating minimisation (AM) procedure, which is split into three steps:

**Step 1:** Populate missing k-space by denoising $\mathbf{x}$ in intermediate domain $(\Phi_F \mathbf{x})_i$,

$$G_P := \min_{\mathbf{x}} \sum_i G((\Phi_F \mathbf{x})_i, \mathbf{y}_i | \Theta_P). \quad (14)$$

**Step 2:** Using this upgraded estimate of k-space we denoise columns $\mathbf{x}_i$ by solving,

$$G_F := \min_{\mathbf{x}} \sum_i G(\mathbf{x}_i, (\Phi_F^H \mathbf{y})_i | \Theta_F). \quad (15)$$

Steps 1 and 2 can be solved in the PG technique described earlier in this section and illustrated in Figures 3a, 3b. The steps will be defined as solving $G_P$ and $G_F$ and their iterations are $n_p$ and $n_f$ respectively. Weights can be shared between successive $G_P$ or $G_F$ iterations but not between $G_P$ and $G_F$ themselves. This ensures AliasNet learns the appropriate intermediate Fourier or image domain denoisers. Figure 3d visualises the reconstruction performed by $G_P$ and $G_F$, and how they are cascaded into one another.

**Step 3:** The 2D regularisation can be any existing DL model by setting $\Psi_{2D}(\cdot)$ to the appropriate regulariser. For instance, regularisation in the denoising sense as per DcCNN [37] (and MoDL [40]),

$$\Psi_{2D}(\mathbf{x}) = \|\mathbf{x} - f_{cnn}(\mathbf{x}|\theta_f)\|_2^2, \quad (16)$$

or regularisation by transformation sparsity as per ISTA-Net [39],

$$\Psi_{2D}(\mathbf{x}) = \|\mathcal{R}(\mathbf{x})\|_1. \quad (17)$$

In fact, 2D regularisation can even be that defined by dual-domain algorithms such as DuDoRNet [53]. Figure 3c illustrates our 1D regularisation combined with a general 2D regularisation for reference. Here, $n_d$ are the total number of iterations to solve the AM algorithm described in this section.

### 2.3.2 Loss Function

While the residual between image domain, intermediate k-space domains and k-space are linearly related, we found that gradient propagation to the 1D $G_F$ and $G_P$ layers is improved by directly supervising the intermediate domains they operate on. As these operations occur on decoupled representations of the image, it suggests the following:

- $G_F$: The forward pass of a single image column $x_i$ through $G_F$ affects only itself and its correlated intermediate Fourier column $(\Phi_F^H y)_i$, whilst affecting all columns of intermediate Fourier domain $\Phi_F x$ and k-space $y$.
- $G_P$: The forward pass of a single intermediate Fourier column $(\Phi_F x)_i$ through $G_P$ affects all columns of the image $x$ and intermediate Fourier domain $\Phi_F^H y$, whilst only itself and correlated columns of k-space $y_i$ are changed.

We, therefore, propose a loss function that resembles the combination of losses required to train $G_F$ and $G_P$ independently of the entire image:

$$L = \|x_0 - \hat{x}\|_1 + \tau_k \|y_0 - \hat{y}\|_1 + \tau_p \|\Phi_P^H y_0 - \Phi_P^H \hat{y}\|_1 + \tau_f \|\Phi_F^H y_0 - \Phi_F^H \hat{y}\|_1. \qquad (18)$$

Here, $x_0$ and $y_0$ are the target image and k-space measurements, $\hat{x}$ and $\hat{y}$ are the estimated image and k-space. Additionally, $\tau_k, \tau_p, \tau_f$ are the k-space, intermediate $\Phi_F$ and $\Phi_P$ domain loss ratios. We found that the approach was relatively insensitive to hyperparameter choice, with the image domain loss generally providing sufficient supervision. Therefore, we set each to 0.1, 0.3, and 0.3 to focus the training on image features, and equally weight the recovery of intermediate k-space in the aliased phase-encoding direction ($\Phi_P^H y$) and zero-filled frequency-encoding direction ($\Phi_F^H y$). Compared to mean absolute error (MAE) alone, convergence characteristics at later epochs are slightly improved, with an accompanying improvement to image quality for our AliasNet models (approximately 0.1dB on average). These findings indicate that Eq. (18) helps to train the decoupled system by directly supervising the domains $G_F$ and $G_P$ operate on, whilst enabling them to learn a regularisation that doesn't negatively impact "unseen" image regions. Furthermore, use of a multi-domain loss function is consistent with DuDoRNet, which due to its dual-domain architecture, deploys a sophisticated multi-layer k-space and image domain loss. While a simple MSE or MAE is sufficient to train our model, the proposed loss function [Eq. (18)], was useful in our comparisons and helped AliasNet to surpass or match DuDoRNet in terms of peak signal-to-noise ratio (PSNR) and structural similarity (SSIM).

# 3 RESULTS

## 3.1 EXPERIMENTAL CONFIGURATION

Two complex valued public MR datasets have been used to train and test our proposed method, each featuring raw 2D k-space of three-dimensional (3D) volumes. Image domain magnitudes of all volumes were normalised between $[0, 1]$. There was no overlap between training, validation, or test sets. Single-coil images are obtained with emulated single-coil methodology as-per each dataset's implementation [58,59]. As the intent of AliasNet is to improve the regularisation of phase-encode artefacts, single-coil images are utilised in this study to observe the improved regularisation of image artefacts without additional spatial correlations. Further to this, the tested 2D and dual-domain networks were originally developed for single-coil image reconstruction [37,39,53,54].

### 3.1.1 Calgary-Campinas

We use the Calgary-Campinas brain dataset [58] to investigate the behaviour of our proposed 1D regularisation technique in several ablation studies. The dataset consists of 45 fully sampled T1w volumes acquired from a 12-channel clinical MR scanner (Discovery MR750; General Electric (GE) Healthcare, Waukesha, WI). In total, there are 7654 slices. Training consists of 25 subjects and 4254 slices. Validation consists of 10 subjects and 1700 slices. To avoid evaluating metrics over background images, the test set consists of 10 subjects and the central 1270 slices. Matrix size is $256 \times 256$.

### 3.1.2 FastMRI

We also train and test on a subset of the NYU fastMRI single-coil knee dataset [59,60]. In this study, we utilise the 3 Tesla coronal proton density (PD)-weighted images without fat suppression. This dataset consists of 12,366 slices from 332 subjects, captured on one of three clinical 3T MR scanners (Siemens Magnetom Skyra, Prisma and Biograph mMR). The data was acquired from a 15-channel knee coil array. Training consists of 8293 slices from 223 volumes. The validation set contains 2081 slices from 56 volumes. The test set is 1567 central slices from 53 volumes. Matrix size is $320 \times 320$.

### 3.1.3 Training Methodology

Discrete Fourier space was sampled using fixed 1D Gaussian random masks for each under-sampling ratio and anatomy. We use PSNR and SSIM to evaluate closeness to the original image. All experiments were conducted on an NVIDIA SXM-2T V100 graphics processing unit (GPU) with 32GB of vRAM. All networks were implemented in PyTorch and trained using the Adam optimiser. All DcCNN and ISTA-Net implementations were trained with a learning rate of $10^{-4}$, a batch size of 10 and with custom loss function [see Eq. (18)]; ISTA-Net was configured as ISTA-Net+ and also had matrix inverse loss as-per [39]. DuDoRNet and MD-Recon-Net were trained following the original implementations [53,54], with batch sizes of 5 and 10 respectively. However, when training on the knee dataset we reduced the number of dilated residual dense blocks (DRD) of DuDoRNet from 3 to 2 within each of its dilated residual dense network (DRD-Net) layers. This is due to the knee dataset being substantially larger than the brain dataset, where 3 DRD will require several weeks to complete just 200 epochs. In contrast, it takes 1 day for our proposed DcCNN + AliasNet network to converge. Additionally, our network features approximately $3 \times$ fewer parameters compared to DuDoRNet, meaning the reduced model is a closer comparison. All networks were trained until convergence, where the network with the lowest validation loss was chosen for testing. Complex numbers are treated as separate input channels by the proposed reconstruction models.

## 3.2 ABLATION STUDIES

To establish a relationship between iterations of $G_P$ and $G_F$ to reconstruction quality, we explore the combination of our proposed 1D AliasNet modules with the 2D regularisation found in DcCNN [37] and ISTA-Net [39]. As-per the AM algorithm described in Section 2.3.1 and Figure 3c, the 1D recovery is interleaved between 2D steps. We also experiment with shared and un-shared configurations. Here, shared 1D modules indicate that the weights of iterations internal to $G_P$ and $G_F$ are common in-between 2D update steps; $G_P$ and $G_F$ do not share weights themselves. Alternatively, un-shared means that every 1D iteration is unique throughout the reconstruction.

### 3.2.1 Impact of AliasNet Iterations

Table 1 compares the total number of unique 1D parameters necessary for shared and un-shared configurations given iterations of $G_P$ ($n_p$), $G_F$ ($n_f$) and 2D steps ($n_d$). Figure 4 compares the relative scaling between iterations of $G_P$ and $G_F$ in both shared (top) and un-shared (bottom) configurations, combined with a reference DcCNN [37] network. DcCNN is configured with 5 data consistency layers ($n_d = 5$), 5 convolutional layers and 32 filters per convolution (D5C5). Experiments are conducted at $R = 4$.

We see that $G_F$ improves the D5C5 estimate, with PSNR converging at $n_f = 5$. By comparison, $G_P$ does not improve PSNR as significantly. This suggests that convolutions applied in image space are more effective than the intermediate domain denoising of $G_P$. We also note that $G_F$ in shared configuration provides similar PSNR scores to un-shared. On the other hand, shared $G_P$ layers do not benefit from $n_p > 1$. For simplicity, all subsequent experiments will cascade $G_P$ and $G_F$ layers by fixing $n_p = 1$ and $n_f = 5$ as-per the configuration in Table 1.

### 3.2.2 Comparison to Large Versions

Using the $n_p = 1, n_f = 5$ configuration, we compare the performance of DcCNN [37] and ISTA-Net [39] boosted with AliasNet layers against larger versions of the original networks. The objective is twofold:

1. Demonstrate that AliasNet enables better scaling with total number of parameters compared to simply increasing the size of the original network.
2. Demonstrate that a combined 1D + 2D CNN improves the regularisation of phase-encode artefacts compared to a 2D only approach.

Average reconstruction performance is summarised in Table 2, where the first rows of DcCNN and ISTA-Net are the baseline models (not boosted by AliasNet). Subsequent rows then increase network size by either increasing the number of filters, increasing the number of cascaded convolutions and in the instance of DcCNN, increasing the number of convolutional layers in each 2D denoising step. AliasNet models are equivalent to the 2D baseline networks, with 1D regularisation interleaved between 2D denoising steps. While the shared configuration uses the same $n_p + n_f = 6$ 1D modules between each 2D denoising step, the un-shared networks deploy a total of $(n_p + n_f) \times n_d = 30$.

For both DcCNN and ISTA-Net and at all reduction factors, the inclusion of AliasNet as an additional 1D regularisation achieves the best PSNR and SSIM scores. This feat is particularly notable for the shared configuration, where only 3.6% and 2.8% of additional parameters are necessary for DcCNN and ISTA-Net respectively. Figures 5a, 5b illustrate the benefits afforded by AliasNet for two sample brain images. It is noted in the error maps that noise-like and faint image features are easily lost by the under-sampling, most evidently at R4. As highlighted in the locations of red and blue arrows, we see AliasNet models better preserve these regions in the reconstruction.

## 3.3 Vs State-of-the-art Algorithms

To further explore AliasNet, we compare against state-of-the-art dual-domain reconstruction techniques DuDoRNet [53] and MD-Recon-Net [54] using the Calgary-Campinas brain and fastMRI knee datasets. As with the ablation study, baseline DcCNN and ISTA-Net networks are also included.

### 3.3.1 Calgary-Campinas Brain

Figures 6a, 6b compare the reconstruction methods for sample images at R4 and R6 of the Calgary-Campinas brain dataset. AliasNet boosted DcCNN and ISTA-Net yield the highest PSNR and SSIM scores. Compared to the state-of-the-art method DuDoRNet, AliasNet improves the reconstruction of the R4 test image from 35.6dB to 36.7dB. The error map image reflects this PSNR advantage. Our findings are also observed in Table 3, which contains each network's average performance at multiple acceleration factors. Surprisingly, adding just 3.6% and 2.8% AliasNet parameters to DcCNN and ISTA-Net respectively, the networks can outperform MD-Recon-Net and DuDoRNet. Further, at higher reduction factors (R6 and R8), MD-Recon-Net cannot improve the reconstruction compared to the baseline DcCNN and ISTA-Net. This is due to large distances between sampled points in k-space limiting the utility of convolutions. DuDoRNet on the other hand retains a performance advantage throughout due to the large receptive field within its DRD-Net modules. AliasNet models are best overall.

### 3.3.2 FastMRI Knee

For the knee images, the frequency and phase-encode directions are transposed to columns and rows respectively. Sample reconstructions at R6 and R8 are demonstrated in Figures 7c, 7d, with average performance over multiple reduction factors presented in Table 3. We see again that AliasNet boosted networks outperform the baseline DcCNN and ISTA-Net implementations. In fact, we also see that the PSNR and SSIM scores have become competitive with DuDoRNet. This is an important outcome, as it is believed that the large receptive field afforded by DuDoRNet's DRD-Net layers is beneficial to the reconstruction of higher resolution knee images ($320 \times 320$) compared to that of the brains ($256 \times 256$). DcCNN and ISTA-Net deploy simple $3 \times 3$ convolutions that struggle to capture long distance interactions by comparison. However, DuDoRNet required approximately 41 days to complete 130 epochs on the 8293 knee slices. As both DcCNN and ISTA-Net employ relatively simple network architectures, and the proposed AliasNet modules are designed to supplement the 2D reconstruction, the AliasNet boosted networks require just 23 hours for convergence. This suggests that our approach to regularise phase-encode under-sampling artefacts in the direction of under-sampling provides a more efficient model from which to recover an MR image.

### 3.3.3 1D Only Reconstruction

Recently, Wang et al. explored a fully 1D CNN to recover multi-coil phase-encode under-sampled MRI [57]. The network achieves state-of-the-art performance when the number of training samples is limited, with performance that remains competitive to 2D techniques as the training samples are increased. The problem posed is similar to that solved in our $G_F$ [see Eq. (15)], with an additional low-rank constraint to interpolate missing k-space from captured values. Our approach differs from this in three key forms. Firstly, their low-rank constraint is applied directly to $\Phi_F^H \boldsymbol{y}$, which we only utilise for data consistency in $G(\boldsymbol{x}_i, (\Phi_F^H \boldsymbol{y})_i | \Theta_F)$ ($G_F$). Secondly, they do not consider 1D aliasing artefacts present in $\Phi_F \boldsymbol{x}$ that we denoise in $G((\Phi_F \boldsymbol{x})_i, \boldsymbol{y}_i | \Theta_P)$ ($G_P$). Lastly, AliasNet was developed to supplement existing DL 2D solvers and therefore, features a minimal number of parameters. We,

therefore, investigate the suitability of AliasNet modules for 1D-only reconstruction. In this configuration, the 1D layers illustrated in Figures 3a, 3b have been scaled into a 687.2K parameter, 1D-only network. Here the number of 1D convolution layers per iteration has increased from 3 to 6, the number of filters per layer from 8 to 32, and $n_p$ and $n_f$ have been set to 5 and 13. The number of parameters is similar to [57]. For simplicity, we continue to experiment with single-coil MRI; AliasNet is easily extended to multi-coil by combining each reconstructed image with the square root of the sum of squares. Figure 8 compares the AliasNet 1D reconstruction, the 2D reconstruction from DcCNN configured in a larger D5C7 and a combined 1D + 2D reconstruction (2D network is D5C5). We find competitive reconstruction performance between 1D and 2D techniques, with PSNR scores only 0.4dB apart. The combined reconstruction however is 1.2dB higher. As per the difference image and the regions indicated by the red and blue arrows, the combined reconstruction better captures image features that are otherwise lost or corrupted in the alternative reconstructions. These findings are supported by Table 4.

We suggest that extension of AliasNet to multi-coil MRI should include the low-rank constraint, such that the reconstruction benefits from standard 2D regularisation, as-well-as 1D regularisation of columns of image $x$ ($G_F$), our proposed 1D regularisation of columns of intermediate domain $\Phi_F x$ ($G_P$) and the low-rank constraint on columns of intermediate domain $\Phi_P x$ as suggested by Wang et al. [57]

## 4 DISCUSSION

To explore the impact of AliasNet modules, we evaluated their performance by pairing with two popular CS-MRI networks. For the sake of fairness, we also examined the effect of increasing the number of 2D operations by either increasing number of filters, increasing number of convolutions, and increasing the number of denoising steps. Our findings suggest that the inclusion of 1D AliasNet layers is more beneficial to the reconstruction compared to additional 2D operations. We then compared the AliasNet boosted networks against dual-domain reconstruction methods and found AliasNet performance was better whilst also requiring fewer parameters and less training time.

Many existing DL CS-MRI algorithms are dependent on image domain operations, only integrating k-space information into data consistency layers and loss function design. Two major limitations have been identified with this approach. Firstly, under-sampling artefacts are structured and non-local, obscuring image features in potentially unrecoverable ways [53]. Secondly, the reconstruction process typically utilises 2D operations, therefore limiting regularisation to an idealised 2D transformation. Recent state-of-the-art contributions have explored dual-domain architectures, executing convolutional operations directly in k-space. However, large distances between sampled points may limit the utility of k-space convolutions. We found that at higher reduction factors for the brain dataset (256 × 256), and all reduction factors for the larger knee images (320 × 320), that MD-Recon-Net [54] is unable to improve the reconstruction compared to DcCNN [37] or ISTA-Net [39]. By comparison, DuDoRNet solves this non-local problem by invoking a sophisticated DRD-Net with large receptive field. While this approach is effective compared to MD-Recon-Net (as demonstrated in our comparisons), the design is computationally expensive. Further it does not directly address the structured nature of phase-encode CS-MRI artefacts. Our proposed AliasNet decouples CS-MRI artefacts into contiguous 1D signals. Further, convolution operations are performed on either the aliased image columns ($G_F$), or aliased intermediate k-space ($G_P$). In either case, convolution in under-sampled k-space is avoided and the expected structure of CS artefacts is directly penalised. This enables our simple AliasNet architecture to improve the reconstruction of DcCNN and ISTA-Net to above DuDoRNet even at high reduction factors.

## 5 CONCLUSION

In this work, we have introduced a novel regularisation layer for phase-encode under-sampled MRI that enhances the reconstruction ability of existing 2D reconstruction techniques. This proposed AliasNet leverages the excellent 1D incoherence properties of phase-encode under-sampled MRI. The combined 1D and 2D regularisation better models image artefacts, producing superior performance scaling compared to simply increasing the size and number of 2D layers in the original networks. In fact, increasing the model size by just 3-4% improves the reconstruction of DcCNN [37] and ISTA-Net [39] to above state-of-the-art dual-domain networks such as MD-Recon-Net [54] and DuDoRNet [53]. We have also found competitive performance from a completely 1D reconstruction network based on AliasNet, potentially enabling high-quality reconstruction for data-constrained applications. In our experiments, image quality benefited more from the proposed image-domain regularisation of $G_F$ [Eq. (15)] than intermediate Fourier regularisation of $G_P$ [Eq. (14)]. However, their combination serves to boost image quality further.

While the objective of this work is to improve existing reconstruction techniques, we suggest a joint 1D and 2D reconstruction strategy be developed that limits the necessary data consistency (DC) layers within each denoising iteration. In the implementation investigated by this paper, we require a 1D DC operation for each 1D AliasNet module. This is a consequence of the AM approach to joint 1D and 2D reconstruction. While this solution yields simple construction and compatibility with existing 2D networks, it is inefficient to perform many fast Fourier transform (FFT) operations. Further, the evaluation of 1D and 2D layers must be performed sequentially rather than in parallel. As such, it may be useful to select and un-roll an optimisation method that limits the number of DC operations and enables parallel computation of regularisation functions.

Another avenue to be explored is the 1D reconstruction of under-sampled projection data, wherein missing sinogram projections are populated via an approach similar to our proposed intermediate Fourier $G_P$.

# 6 REFERENCES

[1] E.J. Candes, J. Romberg, T. Tao, Robust uncertainty principles: exact signal reconstruction from highly incomplete frequency information, IEEE Trans. Inf. Theory. 52 (2006) 489–509. https://doi.org/10.1109/TIT.2005.862083.

[2] D.L. Donoho, Compressed sensing, IEEE Trans. Inf. Theory. 52 (2006) 1289–1306. https://doi.org/10.1109/TIT.2006.871582.

[3] M.F. Duarte, Y.C. Eldar, Structured Compressed Sensing: From Theory to Applications, IEEE Trans. Signal Process. 59 (2011) 4053–4085. https://doi.org/10.1109/TSP.2011.2161982.

[4] M. Lustig, D. Donoho, J.M. Pauly, Sparse MRI: The application of compressed sensing for rapid MR imaging, Magn. Reson. Med. 58 (2007) 1182–1195. https://doi.org/10.1002/mrm.21391.

[5] M. Lustig, D.L. Donoho, J.M. Santos, J.M. Pauly, Compressed Sensing MRI, IEEE Signal Process. Mag. 25 (2008) 72–82. https://doi.org/10.1109/MSP.2007.914728.

[6] C.G. Graff, E.Y. Sidky, Compressive sensing in medical imaging, Appl. Opt. 54 (2015) C23. https://doi.org/10.1364/AO.54.000C23.

[7] M. Sandilya, S.R. Nirmala, Compressed sensing trends in magnetic resonance imaging, Eng. Sci. Technol. Int. J. 20 (2017) 1342–1352. https://doi.org/10.1016/j.jestch.2017.07.001.

[8] C. Lazarus, P. Weiss, N. Chauffert, F. Mauconduit, L. El Gueddari, C. Destrieux, I. Zemmoura, A. Vignaud, P. Ciuciu, SPARKLING: variable-density k-space filling curves for accelerated $T_2^*$-weighted MRI, Magn. Reson. Med. 81 (2019) 3643–3661. https://doi.org/10.1002/mrm.27678.

[9] S. Geethanath, R. Reddy, A.S. Konar, S. Imam, R. Sundaresan, R.B. D. R., R. Venkatesan, Compressed Sensing MRI: A Review, Crit. Rev. Biomed. Eng. 41 (2013) 183–204. https://doi.org/10.1615/CritRevBiomedEng.2014008058.

[10] M. Sandilya, S.R. Nirmala, Compressed sensing trends in magnetic resonance imaging, Eng. Sci. Technol. Int. J. 20 (2017) 1342–1352. https://doi.org/10.1016/j.jestch.2017.07.001.

[11] S.S. Chandra, M. Bran Lorenzana, X. Liu, S. Liu, S. Bollmann, S. Crozier, Deep learning in magnetic resonance image reconstruction, J. Med. Imaging Radiat. Oncol. 65 (2021) 564–577. https://doi.org/10.1111/1754-9485.13276.

[12] E.J. Candès, The restricted isometry property and its implications for compressed sensing, Comptes Rendus Math. 346 (2008) 589–592. https://doi.org/10.1016/j.crma.2008.03.014.

[13] B. Deka, S. Datta, CS-MRI Reconstruction Problem, in: Compress. Sens. Magn. Reson. Image Reconstr. Algorithms Convex Optim. Approach, Springer Singapore, Singapore, 2019: pp. 23–29. https://doi.org/10.1007/978-981-13-3597-6_2.

[14] M. Seeger, H. Nickisch, R. Pohmann, B. Schölkopf, Optimization of $k$-space trajectories for compressed sensing by Bayesian experimental design: Bayesian Optimization of $k$-Space Trajectories, Magn. Reson. Med. 63 (2010) 116–126. https://doi.org/10.1002/mrm.22180.

[15] F. Krahmer, R. Ward, Stable and Robust Sampling Strategies for Compressive Imaging, IEEE Trans. Image Process. 23 (2014) 612–622. https://doi.org/10.1109/TIP.2013.2288004.

[16] B. Adcock, A.C. Hansen, C. Poon, B. Roman, BREAKING THE COHERENCE BARRIER: A NEW THEORY FOR COMPRESSED SENSING, Forum Math. Sigma. 5 (2017) e4. https://doi.org/10.1017/fms.2016.32.

[17] Y. Yang, F. Liu, Z. Jin, S. Crozier, Aliasing Artefact Suppression in Compressed Sensing MRI for Random Phase-Encode Undersampling, IEEE Trans. Biomed. Eng. 62 (2015) 2215–2223. https://doi.org/10.1109/TBME.2015.2419372.

[18] S. Ramani, J.A. Fessler, Parallel MR Image Reconstruction Using Augmented Lagrangian Methods, IEEE Trans. Med. Imaging. 30 (2011) 694–706. https://doi.org/10.1109/TMI.2010.2093536.

[19] S. Ravishankar, Y. Bresler, MR Image Reconstruction From Highly Undersampled k-Space Data by Dictionary Learning, IEEE Trans. Med. Imaging. 30 (2011) 1028–1041. https://doi.org/10.1109/TMI.2010.2090538.

[20] S. Ravishankar, Y. Bresler, Sparsifying transform learning for Compressed Sensing MRI, in: 2013 IEEE 10th Int. Symp. Biomed. Imaging, IEEE, San Francisco, CA, USA, 2013: pp. 17–20. https://doi.org/10.1109/ISBI.2013.6556401.

[21] X. Qu, Y. Hou, F. Lam, D. Guo, J. Zhong, Z. Chen, Magnetic resonance image reconstruction from undersampled measurements using a patch-based nonlocal operator, Med. Image Anal. 18 (2014) 843–856. https://doi.org/10.1016/j.media.2013.09.007.

[22] W. Dong, G. Shi, X. Li, Y. Ma, F. Huang, Compressive Sensing via Nonlocal Low-Rank Regularization, IEEE Trans. Image Process. 23 (2014) 3618–3632. https://doi.org/10.1109/TIP.2014.2329449.

[23] Z. Zhan, J.-F. Cai, D. Guo, Y. Liu, Z. Chen, X. Qu, Fast Multiclass Dictionaries Learning With Geometrical Directions in MRI Reconstruction, IEEE Trans. Biomed. Eng. 63 (2016) 1850–1861. https://doi.org/10.1109/TBME.2015.2503756.

[24] S. Wang, J. Liu, X. Peng, P. Dong, Q. Liu, D. Liang, Two-Layer Tight Frame Sparsifying Model for Compressed Sensing Magnetic Resonance Imaging, BioMed Res. Int. 2016 (2016) 1–7. https://doi.org/10.1155/2016/2860643.

[25] B. Wen, S. Ravishankar, L. Pfister, Y. Bresler, Transform Learning for Magnetic Resonance Image Reconstruction: From Model-Based Learning to Building Neural Networks, IEEE Signal Process. Mag. 37 (2020) 41–53. https://doi.org/10.1109/MSP.2019.2951469.

[26] J.P. Haldar, Low-Rank Modeling of Local k-Space Neighborhoods (LORAKS) for Constrained MRI, IEEE Trans. Med. Imaging. 33 (2014) 668–681. https://doi.org/10.1109/TMI.2013.2293974.

[27] X. Zhang, H. Lu, D. Guo, Z. Lai, H. Ye, X. Peng, B. Zhao, X. Qu, Accelerated MRI Reconstruction With Separable and Enhanced Low-Rank Hankel Regularization, IEEE Trans. Med. Imaging. 41 (2022) 2486–2498. https://doi.org/10.1109/TMI.2022.3164472.

[28] K.P. Pruessmann, M. Weiger, M.B. Scheidegger, P. Boesiger, SENSE: Sensitivity encoding for fast MRI, Magn. Reson. Med. 42 (1999) 952–962. https://doi.org/10.1002/(SICI)1522-2594(199911)42:5<952::AID-MRM16>3.0.CO;2-S.

[29] M.A. Griswold, P.M. Jakob, R.M. Heidemann, M. Nittka, V. Jellus, J. Wang, B. Kiefer, A. Haase, Generalized autocalibrating partially parallel acquisitions (GRAPPA), Magn. Reson. Med. 47 (2002) 1202–1210. https://doi.org/10.1002/mrm.10171.

[30] M. Lustig, J.M. Pauly, SPIRiT: Iterative self-consistent parallel imaging reconstruction from arbitrary $k$-space, Magn. Reson. Med. 64 (2010) 457–471. https://doi.org/10.1002/mrm.22428.

[31] S. Wang, S. Tan, Y. Gao, Q. Liu, L. Ying, T. Xiao, Y. Liu, X. Liu, H. Zheng, D. Liang, Learning Joint-Sparse Codes for Calibration-Free Parallel MR Imaging, IEEE Trans. Med. Imaging. 37 (2018) 251–261. https://doi.org/10.1109/TMI.2017.2746086.

[32] S. Wang, T. Xiao, Q. Liu, H. Zheng, Deep learning for fast MR imaging: A review for learning reconstruction from incomplete k-space data, Biomed. Signal Process. Control. 68 (2021) 102579. https://doi.org/10.1016/j.bspc.2021.102579.

[33] S. Wang, Z. Su, L. Ying, X. Peng, S. Zhu, F. Liang, D. Feng, D. Liang, Accelerating magnetic resonance imaging via deep learning, in: 2016 IEEE 13th Int. Symp. Biomed. Imaging ISBI, IEEE, Prague, Czech Republic, 2016: pp. 514–517. https://doi.org/10.1109/ISBI.2016.7493320.

[34] G. Yang, S. Yu, H. Dong, G. Slabaugh, P.L. Dragotti, X. Ye, F. Liu, S. Arridge, J. Keegan, Y. Guo, D. Firmin, DAGAN: Deep De-Aliasing Generative Adversarial Networks for Fast Compressed Sensing MRI Reconstruction, IEEE Trans. Med. Imaging. 37 (2018) 1310–1321. https://doi.org/10.1109/TMI.2017.2785879.

[35] T.M. Quan, T. Nguyen-Duc, W.-K. Jeong, Compressed Sensing MRI Reconstruction Using a Generative Adversarial Network With a Cyclic Loss, IEEE Trans. Med. Imaging. 37 (2018) 1488–1497. https://doi.org/10.1109/TMI.2018.2820120.

[36] M. Mardani, E. Gong, J.Y. Cheng, S.S. Vasanawala, G. Zaharchuk, L. Xing, J.M. Pauly, Deep Generative Adversarial Neural Networks for Compressive Sensing MRI, IEEE Trans. Med. Imaging. 38 (2019) 167–179. https://doi.org/10.1109/TMI.2018.2858752.

[37] J. Schlemper, J. Caballero, J.V. Hajnal, A. Price, D. Rueckert, A Deep Cascade of Convolutional Neural Networks for MR Image Reconstruction, in: M. Niethammer, M. Styner, S. Aylward, H. Zhu, I. Oguz, P.-T. Yap, D. Shen (Eds.), Inf. Process. Med. Imaging, Springer International Publishing, Cham, 2017: pp. 647–658. https://doi.org/10.1007/978-3-319-59050-9_51.

[38] K. Hammernik, T. Klatzer, E. Kobler, M.P. Recht, D.K. Sodickson, T. Pock, F. Knoll, Learning a variational network for reconstruction of accelerated MRI data: Learning a Variational Network for Reconstruction of Accelerated MRI Data, Magn. Reson. Med. 79 (2018) 3055–3071. https://doi.org/10.1002/mrm.26977.

[39] J. Zhang, B. Ghanem, ISTA-Net: Interpretable Optimization-Inspired Deep Network for Image Compressive Sensing, in: Proc. IEEE Conf. Comput. Vis. Pattern Recognit. CVPR, 2018.

[40] H.K. Aggarwal, M.P. Mani, M. Jacob, MoDL: Model-Based Deep Learning Architecture for Inverse Problems, IEEE Trans. Med. Imaging. 38 (2019) 394–405. https://doi.org/10.1109/TMI.2018.2865356.

[41] Y. Yang, J. Sun, H. Li, Z. Xu, ADMM-CSNet: A Deep Learning Approach for Image Compressive Sensing, IEEE Trans. Pattern Anal. Mach. Intell. 42 (2020) 521–538. https://doi.org/10.1109/TPAMI.2018.2883941.

[42] S. Wang, H. Cheng, L. Ying, T. Xiao, Z. Ke, H. Zheng, D. Liang, DeepcomplexMRI: Exploiting deep residual network for fast parallel MR imaging with complex convolution, Magn. Reson. Imaging. 68 (2020) 136–147. https://doi.org/10.1016/j.mri.2020.02.002.

[43] T. Lu, X. Zhang, Y. Huang, D. Guo, F. Huang, Q. Xu, Y. Hu, L. Ou-Yang, J. Lin, Z. Yan, X. Qu, pFISTA-SENSE-ResNet for parallel MRI reconstruction, J. Magn. Reson. 318 (2020) 106790. https://doi.org/10.1016/j.jmr.2020.106790.

[44] B. Zhou, J. Schlemper, N. Dey, S.S. Mohseni Salehi, K. Sheth, C. Liu, J.S. Duncan, M. Sofka, Dual-domain self-supervised learning for accelerated non-Cartesian MRI reconstruction, Med. Image Anal. 81 (2022) 102538. https://doi.org/10.1016/j.media.2022.102538.

[45] M.B. Lorenzana, C. Engstrom, S.S. Chandra, Transformer Compressed Sensing Via Global Image Tokens, in: 2022 IEEE Int. Conf. Image Process. ICIP, IEEE, Bordeaux, France, 2022: pp. 3011–3015. https://doi.org/10.1109/ICIP46576.2022.9897630.

[46] H. Chung, J.C. Ye, Score-based diffusion models for accelerated MRI, Med. Image Anal. 80 (2022) 102479. https://doi.org/10.1016/j.media.2022.102479.

[47] M. Akçakaya, S. Moeller, S. Weingärtner, K. Uğurbil, Scan-specific robust artificial-neural-networks for k-space interpolation (RAKI) reconstruction: Database-free deep learning for fast imaging, Magn. Reson. Med. 81 (2019) 439–453. https://doi.org/10.1002/mrm.27420.

[48] Y. Han, L. Sunwoo, J.C. Ye, k-Space Deep Learning for Accelerated MRI, IEEE Trans. Med. Imaging. 39 (2020) 377–386. https://doi.org/10.1109/TMI.2019.2927101.

[49] B. Zhu, J.Z. Liu, S.F. Cauley, B.R. Rosen, M.S. Rosen, Image reconstruction by domain-transform manifold learning, Nature. 555 (2018) 487–492. https://doi.org/10.1038/nature25988.

[50] T. Eo, H. Shin, Y. Jun, T. Kim, D. Hwang, Accelerating Cartesian MRI by domain-transform manifold learning in phase-encoding direction, Med. Image Anal. 63 (2020) 101689. https://doi.org/10.1016/j.media.2020.101689.

[51] T. Eo, Y. Jun, T. Kim, J. Jang, H. Lee, D. Hwang, KIKI-net: cross-domain convolutional neural networks for reconstructing undersampled magnetic resonance images, Magn. Reson. Med. 80 (2018) 2188–2201. https://doi.org/10.1002/mrm.27201.

[52] R. Souza, R.M. Lebel, R. Frayne, A Hybrid, Dual Domain, Cascade of Convolutional Neural Networks for Magnetic Resonance Image Reconstruction, in: M.J. Cardoso, A. Feragen, B. Glocker, E. Konukoglu, I. Oguz, G. Unal, T. Vercauteren (Eds.), Proc. 2nd Int. Conf. Med. Imaging Deep Learn., PMLR, 2019: pp. 437–446. https://proceedings.mlr.press/v102/souza19a.html.

[53] B. Zhou, S.K. Zhou, DuDoRNet: Learning a Dual-Domain Recurrent Network for Fast MRI Reconstruction with Deep T1 Prior, in: Proc. IEEECVF Conf. Comput. Vis. Pattern Recognit., 2020: pp. 4273–4282.

[54] M. Ran, W. Xia, Y. Huang, Z. Lu, P. Bao, Y. Liu, H. Sun, J. Zhou, Y. Zhang, MD-Recon-Net: A Parallel Dual-Domain Convolutional Neural Network for Compressed Sensing MRI, IEEE Trans. Radiat. Plasma Med. Sci. 5 (2021) 120–135. https://doi.org/10.1109/TRPMS.2020.2991877.

[55] X. Liu, Y. Pang, R. Jin, Y. Liu, Z. Wang, Dual-domain reconstruction network with V-Net and K-Net for fast MRI, Magn. Reson. Med. 88 (2022) 2694–2708. https://doi.org/10.1002/mrm.29400.

[56] Z. Ke, W. Huang, Z.-X. Cui, J. Cheng, S. Jia, H. Wang, X. Liu, H. Zheng, L. Ying, Y. Zhu, D. Liang, Learned Low-Rank Priors in Dynamic MR Imaging, IEEE Trans. Med. Imaging. 40 (2021) 3698–3710. https://doi.org/10.1109/TMI.2021.3096218.

[57] Z. Wang, C. Qian, D. Guo, H. Sun, R. Li, B. Zhao, X. Qu, One-Dimensional Deep Low-Rank and Sparse Network for Accelerated MRI, IEEE Trans. Med. Imaging. 42 (2023) 79–90. https://doi.org/10.1109/TMI.2022.3203312.

[58] R. Souza, O. Lucena, J. Garrafa, D. Gobbi, M. Saluzzi, S. Appenzeller, L. Rittner, R. Frayne, R. Lotufo, An open, multi-vendor, multi-field-strength brain MR dataset and analysis of publicly available skull stripping methods agreement, NeuroImage. 170 (2018) 482–494. https://doi.org/10.1016/j.neuroimage.2017.08.021.

[59] J. Zbontar, F. Knoll, A. Sriram, T. Murrell, Z. Huang, M.J. Muckley, A. Defazio, R. Stern, P. Johnson, M. Bruno, M. Parente, K.J. Geras, J. Katsnelson, H. Chandarana, Z. Zhang, M. Drozdzal, A. Romero, M. Rabbat, P. Vincent, N. Yakubova, J. Pinkerton, D. Wang, E. Owens, C.L. Zitnick, M.P. Recht, D.K. Sodickson, Y.W. Lui, fastMRI: An Open Dataset and Benchmarks for Accelerated MRI, (2019). http://arxiv.org/abs/1811.08839 (accessed March 27, 2023).

[60] F. Knoll, J. Zbontar, A. Sriram, M.J. Muckley, M. Bruno, A. Defazio, M. Parente, K.J. Geras, J. Katsnelson, H. Chandarana, Z. Zhang, M. Drozdzalv, A. Romero, M. Rabbat, P. Vincent, J. Pinkerton, D. Wang, N. Yakubova, E. Owens, C.L. Zitnick, M.P. Recht, D.K. Sodickson, Y.W. Lui, fastMRI: A Publicly Available Raw k-Space and DICOM Dataset of Knee Images for Accelerated MR Image Reconstruction Using Machine Learning, Radiol. Artif. Intell. 2 (2020) e190007. https://doi.org/10.1148/ryai.2020190007.

# 7 TABLES

*Table 1: Number of unique 1D AliasNet layers with respect to 2D update steps $n_d$. Iterations $n_p$ and $n_f$ relate to 1D minimisation problems $G_P$ [Eq. (14)] and $G_F$ [Eq. (15)], respectively. This configuration is used in all AliasNet experiments.*

| | \multicolumn{5}{c}{AliasNet + 5 2D Layers} |
|---|---|---|---|---|---|
| | $n_p$ | $n_f$ | $n_d$ | 1D Layers | 1D Params |
| **Shared** | 1 | 5 | 5 | 6 | 5,298 |
| **Unshared** | | | | 30 | 26,490 |

Table 2: Average reconstruction performance for the Calgary-Campinas brain dataset (mean ± std). Included in this table are the number of parameters for tested networks. **Bold** and underlined indicate best and second-best outcomes. AliasNet models combine $n_p = 1$ ($G_P$), $n_f = 5$ ($G_F$) and $n_d = 5$ (2D denoising steps). AliasNet models with 6 and 30 1D layers are in the shared and unshared configurations respectively. The increase in parameters is with respect to the baseline model without modification (top row of DcCNN and ISTA-Net).

| Method | Conv Layers | DC Layers | 2D Filters | 1D Layers | R=2 PSNR | R=2 SSIM | R=3 PSNR | R=3 SSIM | R=4 PSNR | R=4 SSIM | % Increase Params |
|---|---|---|---|---|---|---|---|---|---|---|---|
| DcCNN | 5 | 5 | 32 | 0 | 43.15 (±1.16) | 0.9855 (±.0023) | 38.37 (±1.36) | 0.9653 (±.0066) | 35.21 (±1.35) | 0.9376 (±.0130) | - |
|  | 5 | 5 | 38 | 0 | 43.25 (±1.16) | 0.9858 (±.0022) | 38.41 (±1.36) | 0.9655 (±.0065) | 35.31 (±1.36) | 0.9387 (±.0129) | 39.9 |
|  | 6 | 5 | 32 | 0 | 43.30 (±1.15) | 0.9859 (±.0022) | 38.54 (±1.35) | 0.9663 (±.0063) | 35.64 (±1.38) | 0.9441 (±.0119) | 32.0 |
|  | 5 | 7 | 32 | 0 | 43.50 (±1.13) | 0.9876 (±.0295) | 39.01 (±1.33) | 0.9696 (±.0056) | 36.01 (±1.39) | 0.9473 (±.0112) | 40.0 |
| AliasNet (DcCNN) | 5 | 5 | 32 | 6 | <u>43.91 (±1.12)</u> | **0.9876 (±.0021)** | <u>39.57 (±1.33)</u> | <u>0.9725 (±.0051)</u> | <u>36.52 (±1.45)</u> | <u>0.9517 (±.0106)</u> | **3.6** |
|  | 5 | 5 | 32 | 30 | **43.94 (±1.11)** | **0.9876 (±.0021)** | **39.68 (±1.31)** | **0.9731 (±.0049)** | **36.72 (±1.43)** | **0.9535 (±.0101)** | <u>18.2</u> |
| ISTA-Net | 6 | 5 | 32 | 0 | 43.23 (±1.15) | 0.9858 (±.0022) | 38.54 (±1.33) | 0.9665 (±.0060) | 35.61 (±1.36) | 0.9426 (±.0117) | - |
|  | 6 | 5 | 38 | 0 | 43.20 (±1.15) | 0.9857 (±.0022) | 38.57 (±1.33) | 0.9665 (±.0060) | 35.58 (±1.36) | 0.9422 (±.0119) | 40.2 |
|  | 6 | 7 | 32 | 0 | 43.54 (±1.13) | 0.9866 (±.0021) | 39.20 (±1.30) | 0.9705 (±.0051) | 36.14 (±1.37) | 0.9485 (±.0106) | 40.0 |
| AliasNet (ISTA-Net) | 5 | 5 | 32 | 6 | <u>43.86 (±1.11)</u> | <u>0.9875 (±.0020)</u> | <u>39.54 (±1.30)</u> | <u>0.9724 (±.0049)</u> | <u>36.50 (±1.43)</u> | <u>0.9517 (±.0102)</u> | **2.8** |
|  | 5 | 5 | 32 | 30 | **43.96 (±1.10)** | **0.9876 (±.0020)** | **39.68 (±1.30)** | **0.9731 (±.0047)** | **36.74 (±1.41)** | **0.9538 (±.0097)** | <u>13.8</u> |

Table 3: Average reconstruction performance for the Calgary-Campinas brain and fastMRI knee datasets (mean ± std). Included are the number of parameters for tested networks and associated training information. **Bold** and underlined indicate best and second-best outcomes respectively. AliasNet models combine $n_p = 1$ ($G_P$), $n_f = 5$ ($G_F$) and $n_d = 5$. 1D layers are in the shared (top) and unshared (bottom) configurations.

| Method | R=2 | | R=4 | | R=6 | | R=8 | | #(K) Param | Mins/ Epoch |
|---|---|---|---|---|---|---|---|---|---|---|
| | PSNR | SSIM | PSNR | SSIM | PSNR | SSIM | PSNR | SSIM | | |
| **Brains** | | | | | | | | | | |
| DcCNN | 43.15 (±1.16) | 0.9855 (±.0023) | 35.21 (±1.35) | 0.9376 (±.0130) | 32.17 (±1.26) | 0.9026 (±.0203) | 29.78 (±1.24) | 0.8560 (±.0307) | **145.6** | **0.9** |
| ISTA-Net | 43.23 (±1.15) | 0.9858 (±.0022) | 35.61 (±1.36) | 0.9426 (±.0117) | 32.43 (±1.26) | 0.9080 (±.0187) | 29.93 (±1.26) | 0.8623 (±.0283) | 191.5 | 1.6 |
| MD-Rec-Net | 43.30 (±1.18) | 0.9860 (±.0023) | 35.43 (±1.48) | 0.9404 (±.0150) | 32.19 (±1.41) | 0.9043 (±.0245) | 29.62 (±1.34) | 0.8542 (±.0347) | 289.3 | 1.3 |
| DuDoRNet | 43.43 (±1.16) | 0.9863 (±.0022) | 35.85 (±1.40) | 0.9467 (±.0113) | 32.86 (±1.32) | 0.9170 (±.0175) | 30.18 (±1.38) | 0.8710 (±.0306) | 649.4 | 13.2 |
| AliasNet (DcCNN) | 43.91 (±1.12) | **0.9876 (±.0021)** | 36.52 (±1.45) | 0.9517 (±.0106) | 32.98 (±1.32) | 0.9177 (±.0180) | 30.27 (±1.29) | 0.8699 (±.0290) | 150.9 | 2.0 |
| | 43.94 (±1.11) | **0.9876 (±.0021)** | 36.72 (±1.43) | 0.9535 (±.0101) | 33.08 (±1.33) | 0.9197 (±.0175) | 30.34 (±1.29) | 0.8721 (±.0288) | 172.1 | 2.1 |
| AliasNet (ISTA-Net) | 43.86 (±1.11) | 0.9875 (±.0020) | 36.50 (±1.43) | 0.9517 (±.0102) | 33.00 (±1.31) | 0.9183 (±.0173) | 30.34 (±1.30) | 0.8718 (±.0283) | 196.8 | 2.5 |
| | **43.96 (±1.10)** | **0.9876 (±.0020)** | **36.74 (±1.41)** | **0.9538 (±.0097)** | **33.13 (±1.31)** | **0.9207 (±.0170)** | **30.37 (±1.32)** | **0.8724 (±.0289)** | 218.0 | 2.6 |
| **Knees** | | | | | | | | | | |
| DcCNN | 35.56 (±3.02) | 0.8833 (±.0626) | 32.36 (±2.73) | 0.7769 (±.0885) | 30.77 (±2.48) | 0.7164 (±.0972) | 29.19 (±2.22) | 0.6622 (±.0976) | **145.6** | **3.6** |
| ISTA-Net | 35.56 (±3.02) | 0.8833 (±.0626) | 32.36 (±2.72) | 0.7778 (±.0881) | 30.83 (±2.48) | 0.7188 (±.0973) | 29.27 (±2.24) | 0.6651 (±.0982) | 191.5 | 6.5 |
| MD-Rec-Net | 35.53 (±3.02) | 0.8831 (±.0625) | 32.34 (±2.72) | 0.7775 (±.0879) | 30.73 (±2.45) | 0.7168 (±.0961) | 29.01 (±2.20) | 0.6603 (±.0965) | 289.3 | 6.1 |
| DuDoRNet | **35.61 (±3.04)** | **0.8842 (±.0628)** | 32.45 (±2.77) | **0.7810 (±.0894)** | 30.98 (±2.53) | **0.7235 (±.0985)** | **29.61 (±2.31)** | **0.6734 (±.1003)** | 452.3 | 465 |
| AliasNet (DcCNN) | 35.58 (±3.03) | 0.8836 (±.0626) | 32.46 (±2.78) | 0.7789 (±.0895) | 30.99 (±2.54) | 0.7208 (±.0986) | 29.48 (±2.30) | 0.6677 (±.0998) | 150.9 | 5.7 |
| | 35.60 (±3.04) | 0.8839 (±.0628) | 32.49 (±2.78) | 0.7797 (±.0894) | 31.04 (±2.56) | 0.7222 (±.0990) | 29.51 (±2.30) | 0.6686 (±.0998) | 172.1 | 6.1 |

| AliasNet | 35.58 (±3.03) | 0.8836 (±.0626) | 32.47 (±2.76) | 0.7796 (±.0891) | 31.01 (±2.54) | 0.7221 (±.0986) | 29.50 (±2.29) | 0.6690 (±.0997) | 196.8 | 9.5 |
| --- | --- | --- | --- | --- | --- | --- | --- | --- | --- | --- |
| (ISTA-Net) | 35.60 (±3.04) | 0.8839 (±.0628) | **32.50 (±2.78)** | 0.7802 (±.0894) | **31.06 (±2.56)** | 0.7233 (±.0990) | **29.61 (±2.31)** | 0.6714 (±.1002) | 218.0 | 9.5 |

Table 4: Comparison between a large 1D only AliasNet, a 2D only reconstruction via DcCNN configured to D5C7 and the combined 1D + 2D reconstruction configured as P1F5 + D5C5. Dataset used is Calgary-Campinas (mean ± std).

| Method | R=4 | | Number of Parameters |
|---|---|---|---|
| | PSNR | SSIM | |
| AliasNet: $n_p$=3, $n_f$=15 | 35.26 (±1.58) | 0.9400 (±.0150) | 687.2K |
| DcCNN: D5C7 | <u>36.01 (±1.39)</u> | <u>0.9473 (±.0112)</u> | <u>203.8K</u> |
| AliasNet (DcCNN: D5C5) | **36.72 (±1.43)** | **0.9535 (±.0101)** | **172.1K** |

# 8 FIGURES AND CAPTIONS

## 8.1 CAPTIONS

## 8.2 Figures

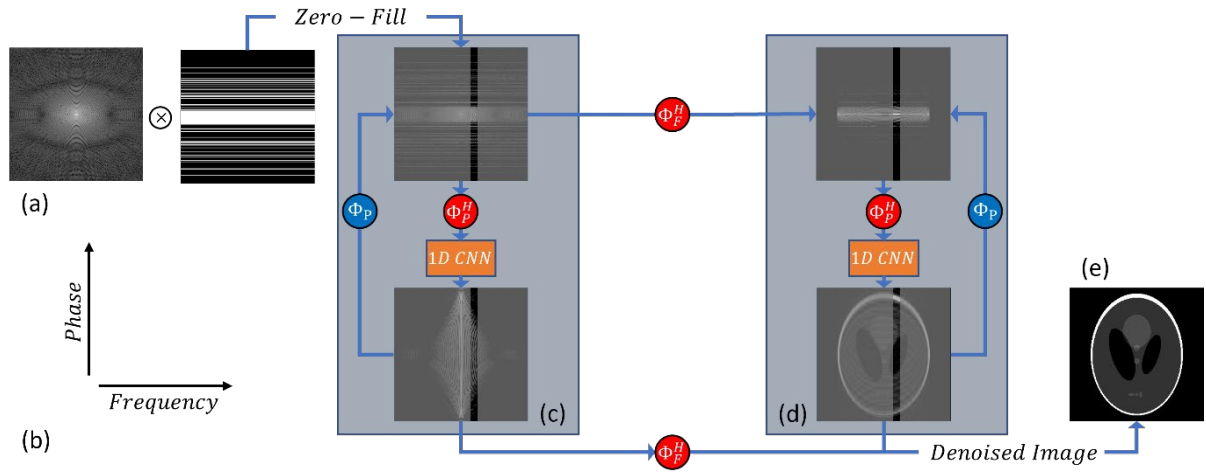

Figure 1: The orange 1D CNN box is our proposed AliasNet module designed to recover columns of aliased 1D Fourier signals. Blue circles indicate the forward 1D DFT and red the inverse, with $\Phi_P$ and $\Phi_F$ specifying the phase- and frequency-encode directions respectively. (a) Is the under-sampling strategy, (b) the phase- and frequency-encode directions, (c) proposed denoising of 1D intermediate Fourier columns, (d) proposed denoising of 1D image columns, (e) the target image.

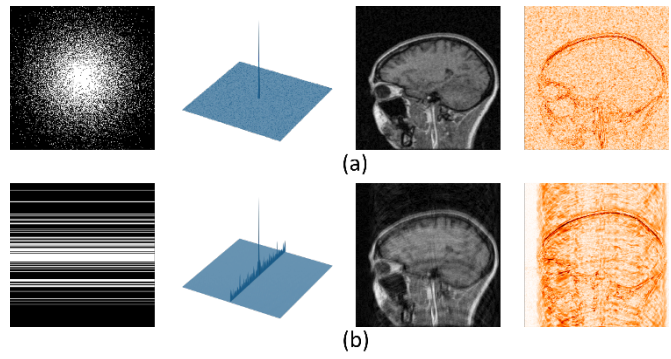

*Figure 2: Visual comparison between artefacts of 2D random (a) and 1D random (b) k-space under-sampling schemes. From left to right are the sampling mask, PSF, zero-filled image and error map.*

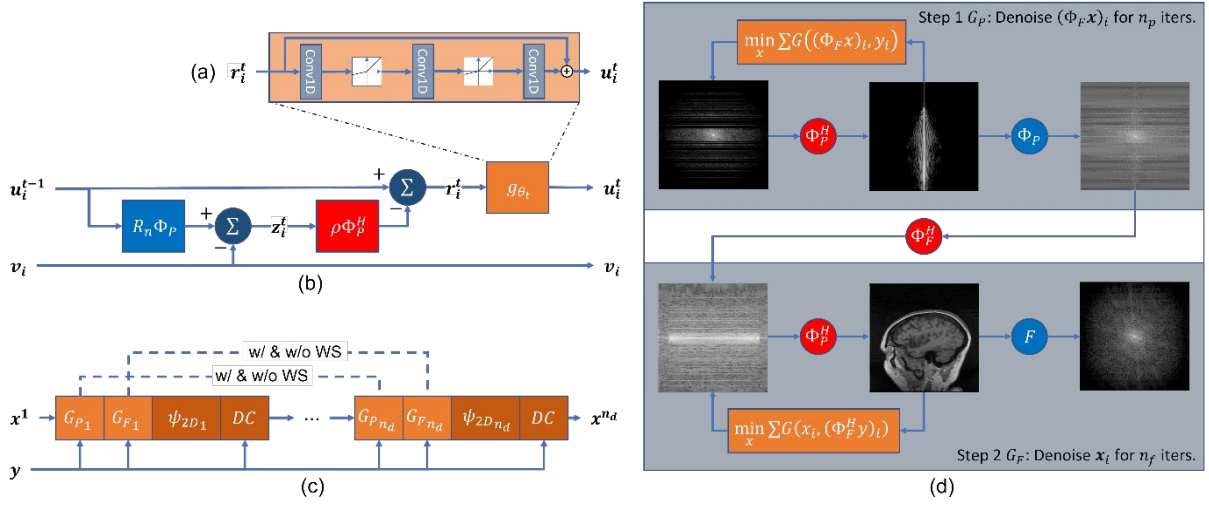

Figure 3: Proposed 1D PG technique for accelerated phase-encode MRI. (a) implementation of 1D denoiser $g_{cnn}(r^t|\theta_t)$, (b) proposed update procedure for $\min_{u} G(u,v|\Theta)$, (c) proposed 1D and 2D reconstruction procedure, where $G_P$ and $G_F$ can be configured to share weights within $G_P$ or $G_F$ iterations (w/ WS), or be independent between each iteration (w/o WS); $n_d$ are the total number of 2D iterations. (d) Illustrated cascade of $G_P$ and $G_F$, which are defined by Eqs. (14, 15).

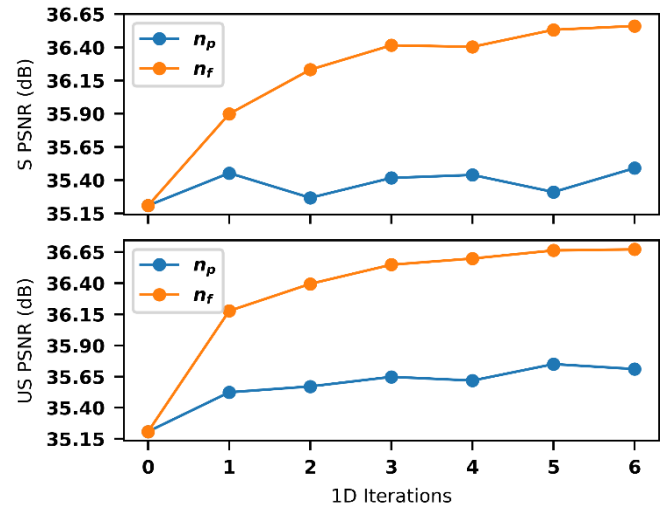

Figure 4: Scaling of the AliasNet layers combined with DcCNN configured to D5C5 ($n_d = 5$). $n_p$ and $n_f$ are iterations of $G_P$ (Eq. 14) and $G_F$ (Eq. 15) respectively. Dataset used is Calgary-Campinas at a reduction factor of 4.

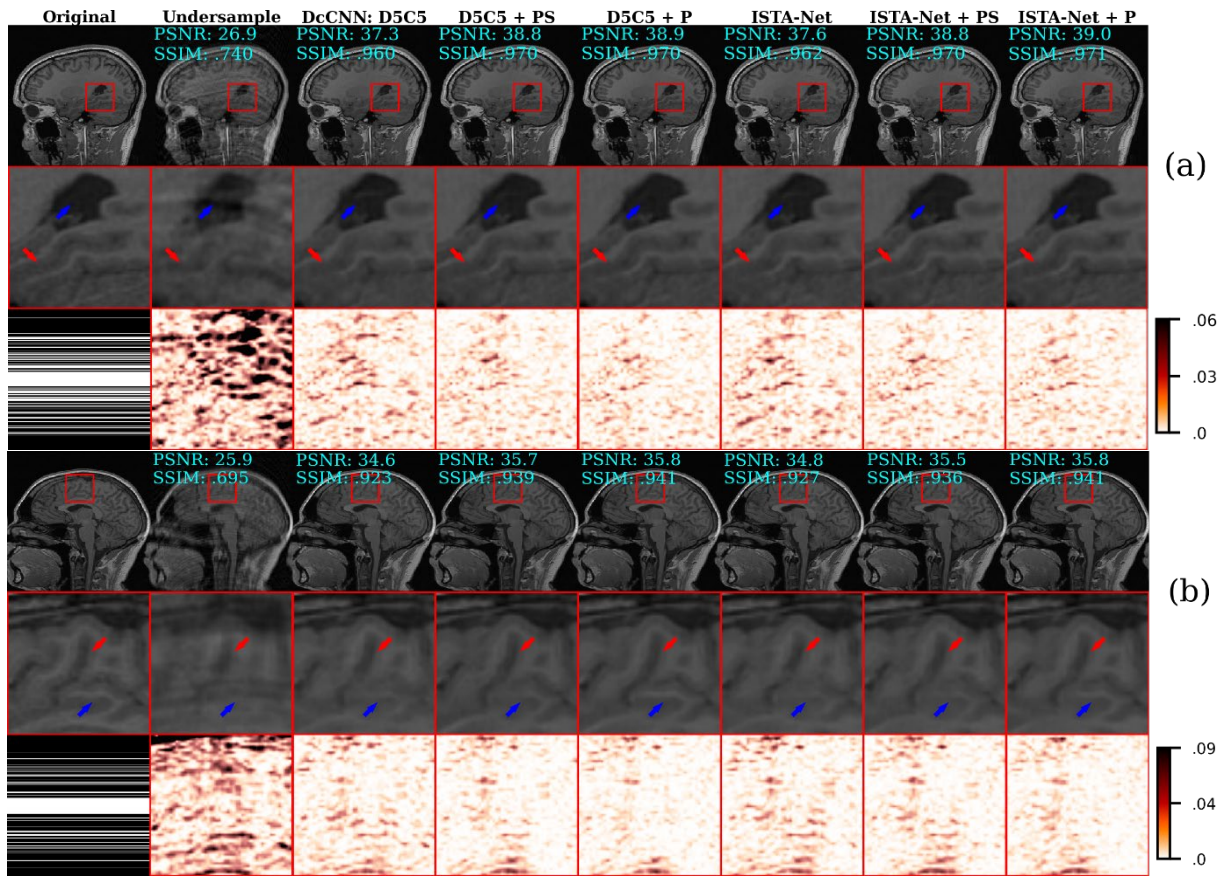

Figure 5: Reconstruction performance demonstrated at R3 (a) and R4 (b) of sagittal cross-sectional brain MR images, with the zoomed in area (red rectangle) highlighting regions where reconstruction performance is notably improved by AliasNet layers. DcCNN and ISTA-Net are the baselines. PS and P indicate the inclusion of AliasNet in shared and un-shared configurations. Values ∈ [0, 1].

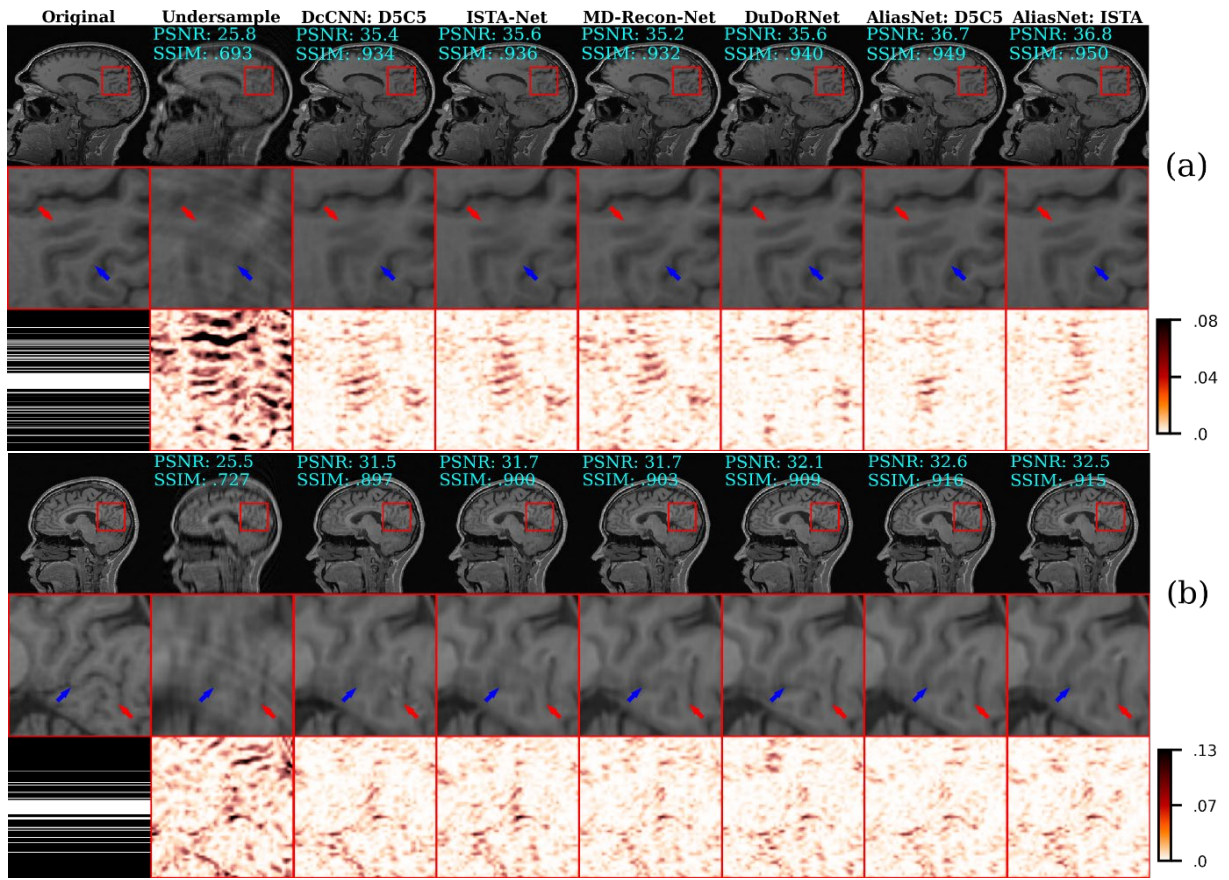

*Figure 6: Reconstruction performance demonstrated for the brains at R4 (a) and R6 (b). Featured in this comparison are DcCNN configured as D5C5, ISTA-Net, dual-domain networks MD-Recon-Net and DuDoRNet, and the AliasNet boosted DcCNN and ISTA-Net implementations without weight sharing. The zoomed area (red rectangle) highlighting regions where the reconstruction is notably improved by AliasNet compared to the baseline DcCNN and ISTA-Net implementations. Values ∈ [0, 1].*

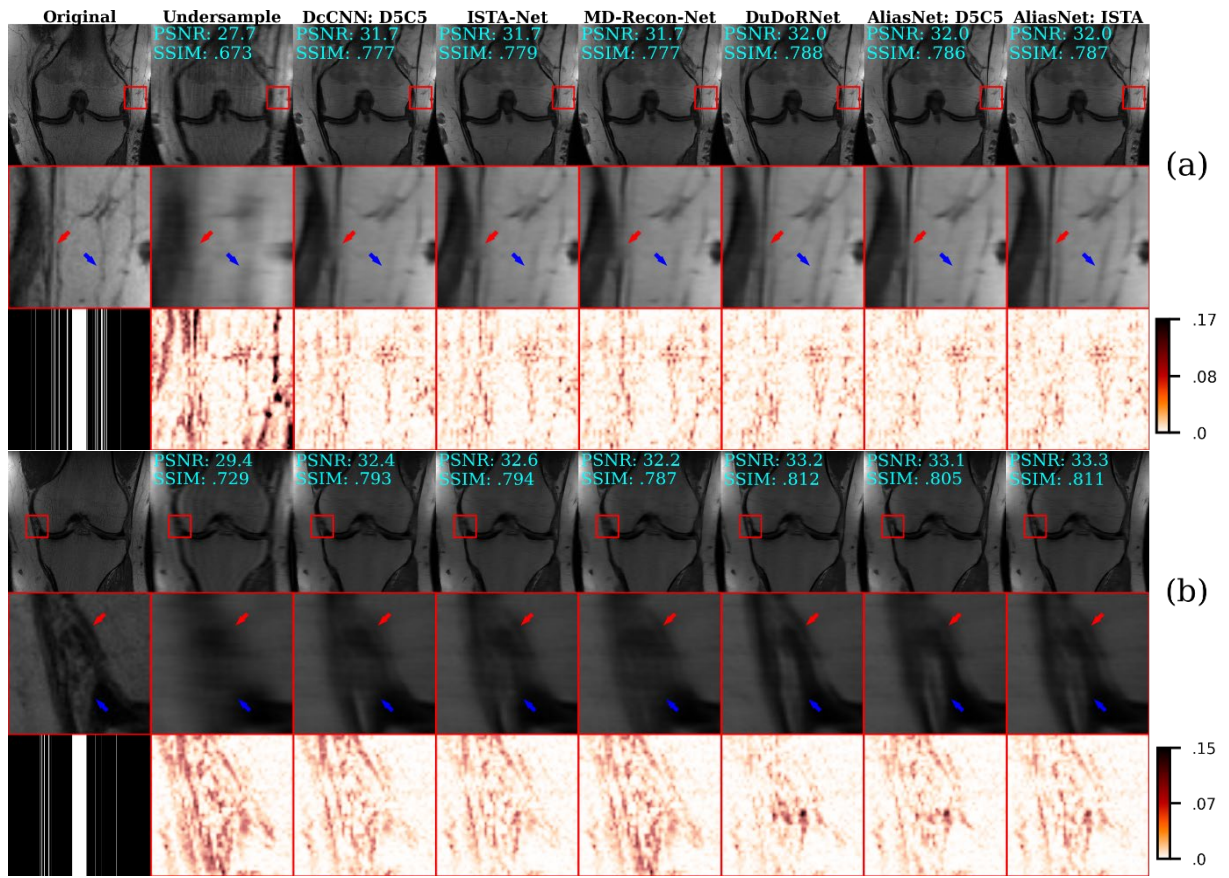

*Figure 7: Reconstruction performance demonstrated for the knees at R6 (a) and R8 (b). Featured in this comparison are DcCNN configured as D5C5, ISTA-Net, dual-domain networks MD-Recon-Net and DuDoRNet, and the AliasNet boosted DcCNN and ISTA-Net implementations without weight sharing. The zoomed area (red rectangle) highlights regions where the reconstruction is notably improved by AliasNet compared to the baseline DcCNN and ISTA-Net implementation. Values ∈ [0, 1].*

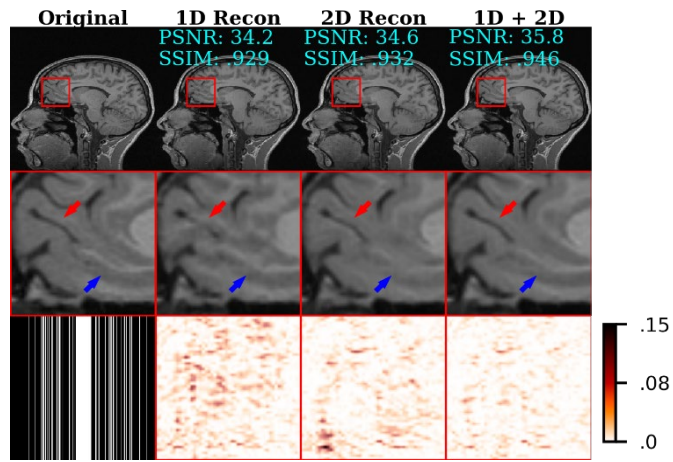

*Figure 8: Comparison of 1D, 2D and combined 1D + 2D reconstructions for the Calgary-Campinas brain dataset at R4. 1D is our AliasNet only, 2D is DcCNN configured to D5C7 and 1D + 2D is AliasNet combined with D5C5.*